\documentclass[preprint]{aastex63}

\shorttitle{Resonant tidal responses}
\shortauthors{Lin \& Ogilvie}

\begin{document}

\title{Resonant tidal responses in rotating fluid bodies: global modes hidden beneath localized wave beams}

\correspondingauthor{Yufeng Lin}
\email{linyf@sustech.edu.cn}

\author[0000-0002-7639-9594]{Yufeng Lin}
\affiliation{Department of Earth and Space Sciences, Southern University of Science and Technology, \\Shenzhen 518055, China}

\author[0000-0002-7756-1944]{Gordon I. Ogilvie}
\affiliation{Department of Applied Mathematics and Theoretical Physics, University of Cambridge, \\ Centre for Mathematical Sciences, Wilberforce Road, Cambridge CB3 0WA, UK}

\begin{abstract}
In rotating stars and planets, excitation of inertial waves in convective envelopes provides an important channel for tidal dissipation, but the dissipation rate due to inertial waves depends erratically on the tidal frequency. Tidal dissipation is significantly enhanced at some frequencies, suggesting possible resonances between the tidal forcing and some eigenmodes. However, the nature of these resonances remains enigmatic owing to the singularity of the eigenvalue problem of inertial waves, and the resonances are often mistakenly attributed to wave attractors in the literature. In this letter, we reveal that resonant tidal responses correspond to inertial modes with large-scale flows hidden beneath localized wave beams. Strong couplings between the tidal forcing and the hidden large-scale flows intensify the localized  wave beams emanating from the critical latitudes, leading to enhanced tidal dissipation. This study resolves a long-standing puzzle regarding the frequency-dependence of tidal dissipation due to inertial waves in convective envelopes.  

\end{abstract}

\section{Introduction} \label{sec:intro}

Tidal interactions play an important role in the spin-orbital evolution of systems involving stars, planets and moons. The problem has been studied for more than a century \citep[e.g.][]{Darwin1880}, yet it remains uncertain to estimate the efficiency of tidal dissipation in real systems. The tidal response is conventionally treated as a hydrostatic deformation, known as the equilibrium tide, which acquires a small phase lag with respect to the tidal forcing because of dissipation \citep{Zahn1966,Zahn1989}. However, the equilibrium tidal response does not satisfy the equation of motion when the tidal frequency is nonzero. Corrections to the equilibrium tide  introduce more complicated dynamical tides, which usually involve internal waves in stars and gaseous planets \citep{Ogilvie2014}. 

A widely studied mechanism for dynamical tides involves internal gravity waves in radiative regions \citep{Zahn1975AA,Savonije1983MNRAS,Goldreich1989ApJ,Goodman1998ApJ,Terquem1998ApJ,Ahuir2021AA}. In rotating stars and planets, inertial waves restored by the Coriolis force can be tidally excited in convective regions when the tidal frequency in the rotating frame is smaller than twice the rotation frequency \citep{Savonije1997MNRAS,Ogilvie2004ApJ,Wu2005ApJ,Wu2005b,Ogilvie2007ApJ,Lai2012MNRAS, Lee2020MNRAS}, providing an additional channel of tidal dissipation. For dynamical tides, the efficiency of tidal dissipation depends strongly on the tidal frequency, as resonances may take place at certain frequencies, leading to enhanced  dissipation \citep{Ogilvie2014}. Furthermore, resonance locking with certain modes has been promoted to explain accelerated tidal evolution in several systems \citep{Witte1999,Burkart2013,Fuller2016MNRAS,Zanazzi2021AJ,Ma2021}. It is therefore of great importance to understand the nature of resonant tidal responses.

The resonance with gravity modes is relatively straightforward, although the rotational and nonlinear effects can complicate the picture \citep{Savonije1997MNRAS,Weinberg2012ApJ}. On the other hand, tidal dissipation due to inertial waves has a more complicated frequency-dependence \citep{Ogilvie2009,Rieutord2010}, probably because of the intrinsic singularity of the inertial wave equations \citep{Stewartson1969,Rieutord1997JFM}.  In a full sphere, smooth inertial modes do exist and analytical solutions can be obtained \citep{Greenspan1968,Zhang2001,Wu2005ApJ}, {although the tidal forcing of these modes vanishes in the case of a uniform density sphere \citep[][but see Wu (2005b) for inhomogeneous models]{Goodman2009ApJ,Ogilvie2013MNRAS}}.  In a spherical shell (i.e. convective envelope), smooth global inertial modes do not exist except purely toroidal modes. Numerical calculations with a small viscosity found that localized wave beams generated by the singularity at a critical latitude on the inner boundary propagate into the bulk along the characteristics of the inertial wave equations \citep{Hollerbach1995JFM,Rieutord1997JFM}, and can form attractors after multiple reflections within certain frequency intervals  \citep{Rieutord2001}. 

The dissipation rate of tidally excited inertial waves  depends erratically on the tidal frequency, and can vary over several orders of magnitude between peaks and troughs  
\citep{Ogilvie2004ApJ,Ogilvie2009,Rieutord2010}. { It is also important to note  that tidal dissipation associated with inertial waves varies over several orders of magnitude along the evolution of stars \citep{Mathis2015AA,Bolmont2016,Barker2020MNRAS}}. Ray dynamics has been used to explain the complicated frequency dependence of the dissipation rate \citep{Rieutord2010,Rieutord2018}, yet several aspects including the resonant peaks remain unexplained. 
 In a  two-dimensional torus model, resonant peaks correspond to attractors or periodic orbits of rays,  which can be described by the ray dynamics \citep{Rieutord2010}. In a real spherical shell, however, neither attractors nor periodic orbits correspond to resonant peaks of the dissipation curves \citep{Ogilvie2009,Rieutord2010,Rekier2019}, though attractors were often misinterpreted as resonant inertial waves in the literature \citep[e.g.][]{Fuller2016MNRAS,Luan2018MNRAS}. Resonant frequencies also depend on the sign of the tidal frequency or azimuthal wavenumber, which cannot be explained by the ray dynamics \citep{Ogilvie2009}.  
 As resonance locking with inertial waves was also promoted to explain the rapid tidal evolution of the satellite systems of giant planets 
\citep{Lainey2020NatAs}, it is critical to understand the nature of these resonances. 
  
In this letter, we use a simplified model introduced in \cite{Ogilvie2009} to investigate resonant tidal responses in rotating fluid bodies, relevant to dynamical tides in convective envelopes of rotating stars and planets. We reveal that large-scale smooth structures, which are reminiscent of global inertial modes in a full sphere,  are hidden beneath the localized wave beams when  resonances take place in a spherical shell. Resonant interactions between the tidal forcing and these large-scale responses intensify the localized wave beams emanating from the critical latitudes, leading to enhanced tidal dissipation. Our results resolve a long-standing puzzle regarding the frequency dependence of tidal dissipation associated with inertial waves in convective envelopes.    

\section{Simplified model} \label{sec:model}
We consider tidal responses in  a uniformly rotating spherical body of radius $r_o$, which  consists of a rigid core of radius $r_i$ and an incompressible homogeneous fluid envelope. As we focus on dynamical tides associated with inertial waves in the fluid envelope, tidal forcing is approximated as a periodic radial flow at the outer boundary in the rotating frame \citep{Ogilvie2009}:
\begin{equation}
u_r=U_0Y_n^m(\theta,\phi) {e}^{- i \omega t},
\end{equation}
where $U_0$ is related to the tidal amplitude, $Y_n^m(\theta,\phi)$ is a standard spherical harmonic, and $\omega$ is the tidal frequency in the rotating frame. Here we use spherical coordinates ($r$, $\theta$, $\phi$). Linear tidal responses in the fluid envelope are governed by the linearized Navier-Stokes equations in the rotating frame:
\begin{equation} \label{eq:NS1}
\frac{\partial \mathbf u}{\partial t}+2\Omega \mathbf{\hat{z} \times u}=-\nabla W+\nu \nabla^2 \mathbf u,
\end{equation} 
\begin{equation}\label{eq:Incompressible}
\mathbf{\nabla \cdot u}=0,
\end{equation} 
where $\mathbf{u}$ is the flow velocity, ${\Omega}$ is the rotation frequency, $\mathbf{\hat z}$ is the unit vector along the rotation axis, $W$ is the reduced pressure including all potential terms, and $\nu$ is the kinematic viscosity. The viscosity is measured by the Ekman number $E=\nu/(\Omega r_o^2)$, i.e. the ratio between the rotation time scale and the global viscous time scale.  We set 
stress-free boundary conditions at both inner and outer boundaries. 

Equations (\ref{eq:NS1} - \ref{eq:Incompressible}) subject to the boundary conditions are solved using a pseudo-spectral method based on an expansion of spherical harmonics on spherical surfaces and Chebyshev collocation in the radial direction. As we consider linear responses to a periodic forcing, the time derivative can be expressed as $\partial \mathbf{u} /\partial t=-i \omega \mathbf u$. We actually solve the boundary value problem in the frequency domain.  The detailed numerical scheme can be found in \cite{Ogilvie2009}. By numerically solving the problem, we calculate the time-averaged kinetic energy $E_k$ and dissipation rate $D_k$ given by:
\begin{equation}
E_k=\frac{1}{4}\rho \int_V \mathbf {u\cdot u^*} \mathrm d V,
\end{equation}
\begin{equation}
D_k=\frac{1}{2}\rho  \int_V \nu \nabla^2 \mathbf {u\cdot u^*} \mathrm d V,
\end{equation}
where $\mathbf {u^*}$ denotes the complex conjugate.

We also solve the unforced eigenmodes of the system  by assuming $\mathbf{u} \propto e^{-i \omega t}$, where $\omega$ is a complex eigenvalue for the unforced problem. The real part and imaginary part of $\omega$ represent the frequency and decay rate of an eigenmode respectively. Using the same method as the forced problem for the  spatial discretization, we end up with a generalized eigenvalue problem, which is solved using an iterative method to find the least damped eigenmode around a given frequency $\omega_0$ \citep{Rieutord1997JFM}.

\begin{figure*}
   \centering
   \includegraphics[width=0.99 \textwidth]{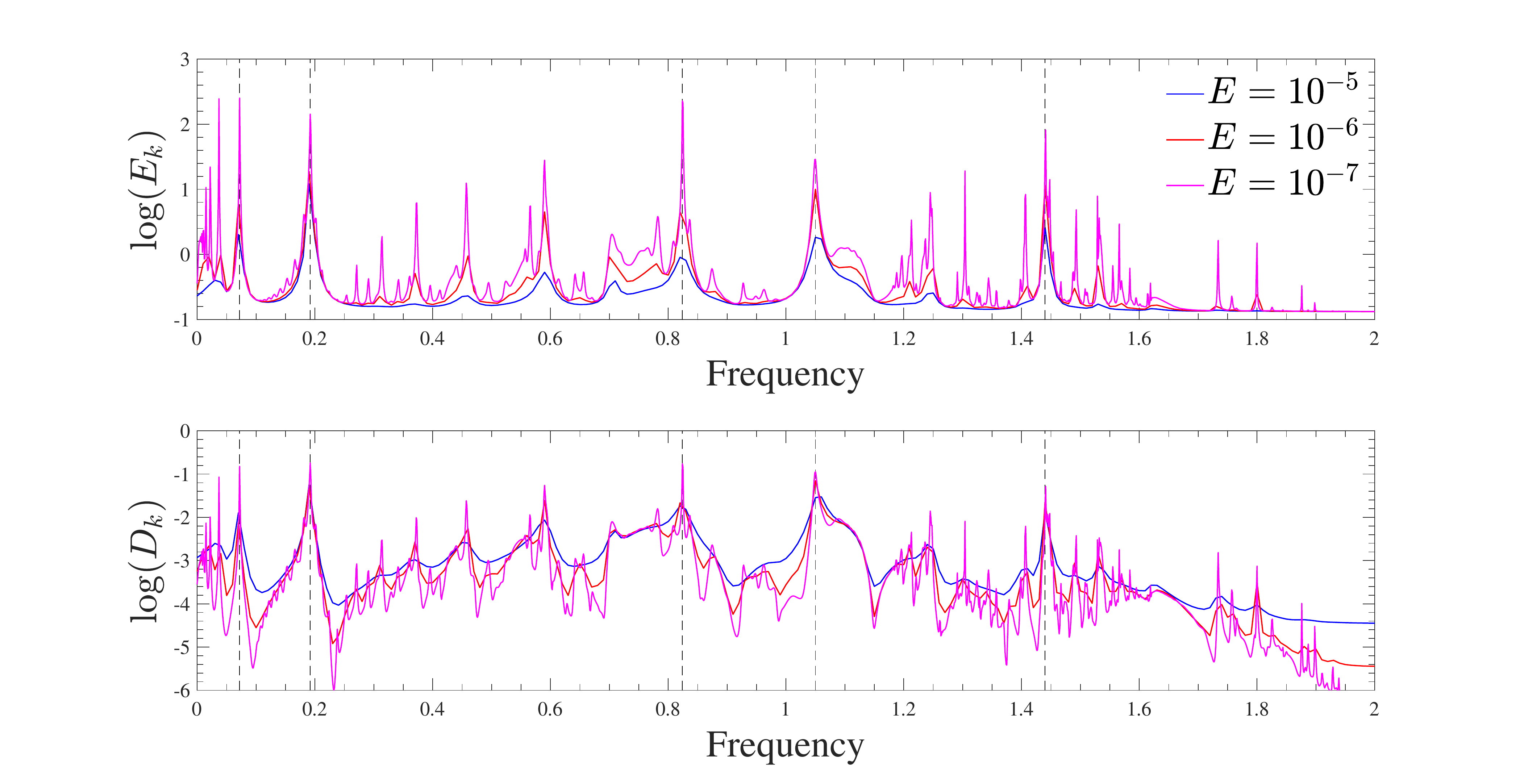}
   \caption{Kinetic energy (in units of $\rho U_0^2r_0^3$) and dissipation rate (in units of $\rho \Omega U_0^2r_0^3$) as a function of the non-dimensional tidal frequency $\omega/\Omega$ for various Ekman numbers.}
    \label{fig:FreScan}%
\end{figure*}

\section{Results} \label{sec:results}
In this letter, we consider only the dominant tidal component $l=2$ and $m=2$. For the $l=2$ and $m=1$ component in spin-orbit misaligned systems, the spin-over mode (purely toroidal mode) is involved and requires special treatment \citep{Lin2017MNRAS}. We choose a moderate inner core size $r_i/r_o=0.5$ for the purposes of illustration. For a very small inner core (i.e. $r_i\ll r_o$) or a large core (i.e. $r_o-r_i\ll r_o$), the problem can be treated using the full sphere model or the shallow water model respectively.

\autoref{fig:FreScan} shows the kinetic energy $E_k$ and dissipation rate $D_k$ as a function of the non-dimensional tidal frequency $\omega/\Omega$. Similar curves have been shown in previous studies \citep{Ogilvie2009,Rieutord2010}, exhibiting erratic dependence on the tidal frequency.  At certain frequencies (e.g. vertical dashed lines in figure \ref{fig:FreScan}),  both the tidal energy and the dissipation are significantly enhanced, and the dissipation rate increases on reducing the viscosity (Ekman number), suggesting possible resonances with eigenmodes at these privileged frequencies. Such a frequency-dependence is reminiscent of the excitation of inertial modes in a full sphere ( e.g. figure 3 in \cite{Aldridge1969} ), but no regular inertial modes generally  exist  in a spherical shell as we have mentioned. Nevertheless, eigenmodes in a spherical shell can be obtained numerically by introducing a small viscosity. A well-known type of eigenmode in a spherical shell is the so-called attractor mode, in which wave beams are focused towards a closed path \citep{Rieutord2001}, but attractors do not correspond to the resonant peaks in the dissipation curves \citep{Ogilvie2009,Rieutord2010}. The nature of these resonant peaks has remained enigmatic.
\begin{figure}
   \centering
   \includegraphics[width=0.45 \textwidth]{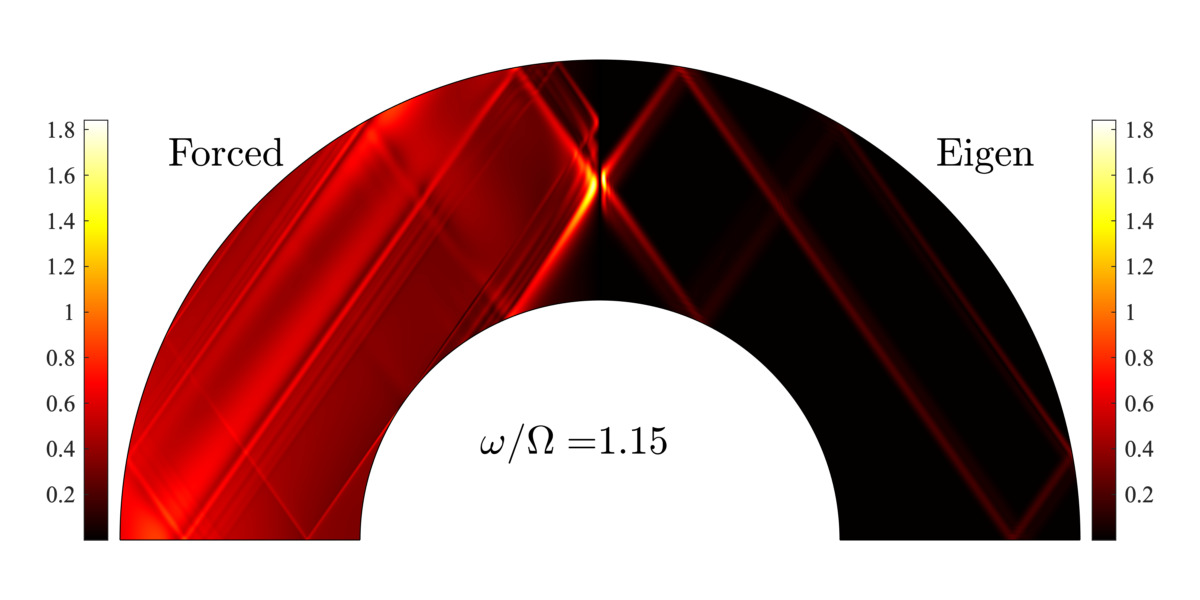} 
\includegraphics[width=0.45 \textwidth]{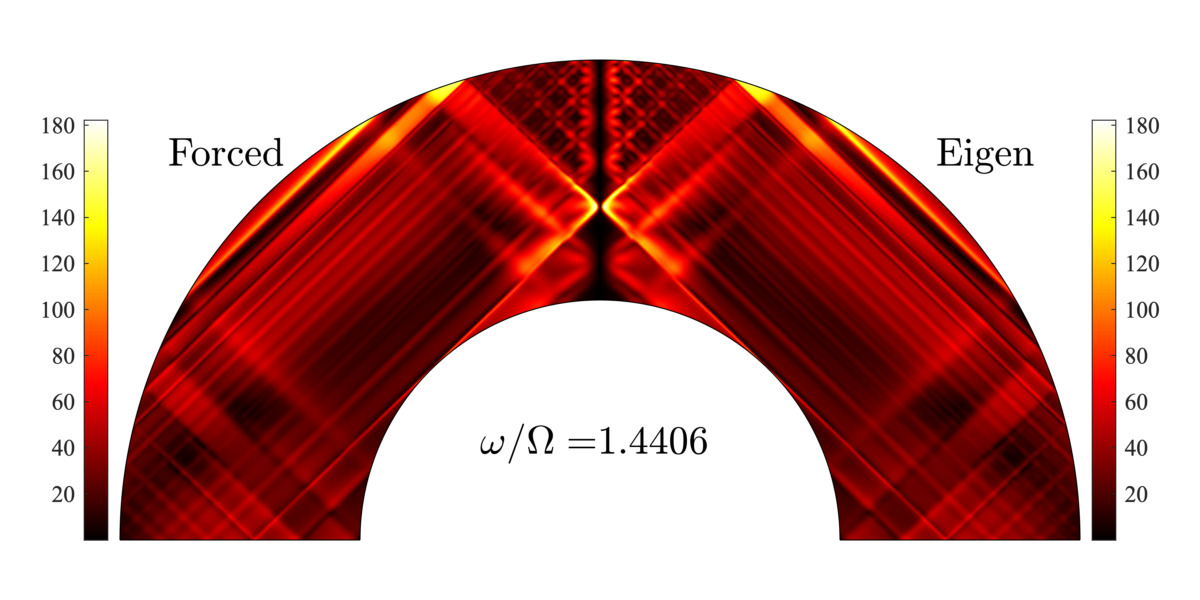} \\
(a) \hspace{7cm} (b)
   \caption{Velocity amplitude $|\mathbf u|$ in the meridional plane (only a quarter is shown due to the symmetry) at (a) $\omega/\Omega=1.15$ (corresponding to a trough)  and (b) $\omega/\Omega=1.4406$ (corresponding to a peak).  Left parts show the tidally forced flows ($U_0=1$) and right parts show the least damped eigenmodes around the forcing frequencies. The Ekman number $E=10^{-8}$ for all cases.}
    \label{fig:ForcedEigen}%
\end{figure}

\begin{figure*}
\centering
\includegraphics[width=0.4 \textwidth]{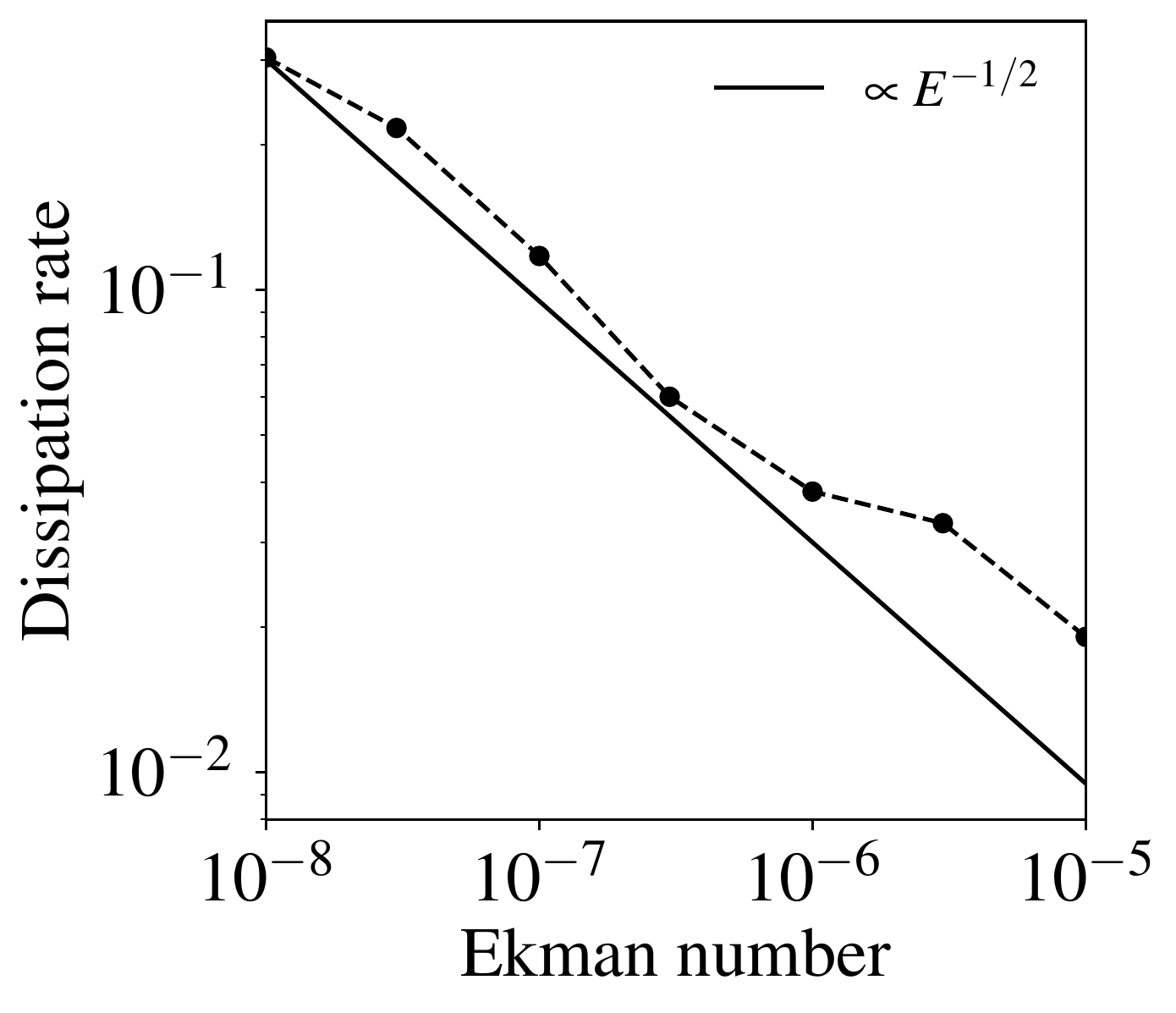}
\includegraphics[width=0.4 \textwidth]{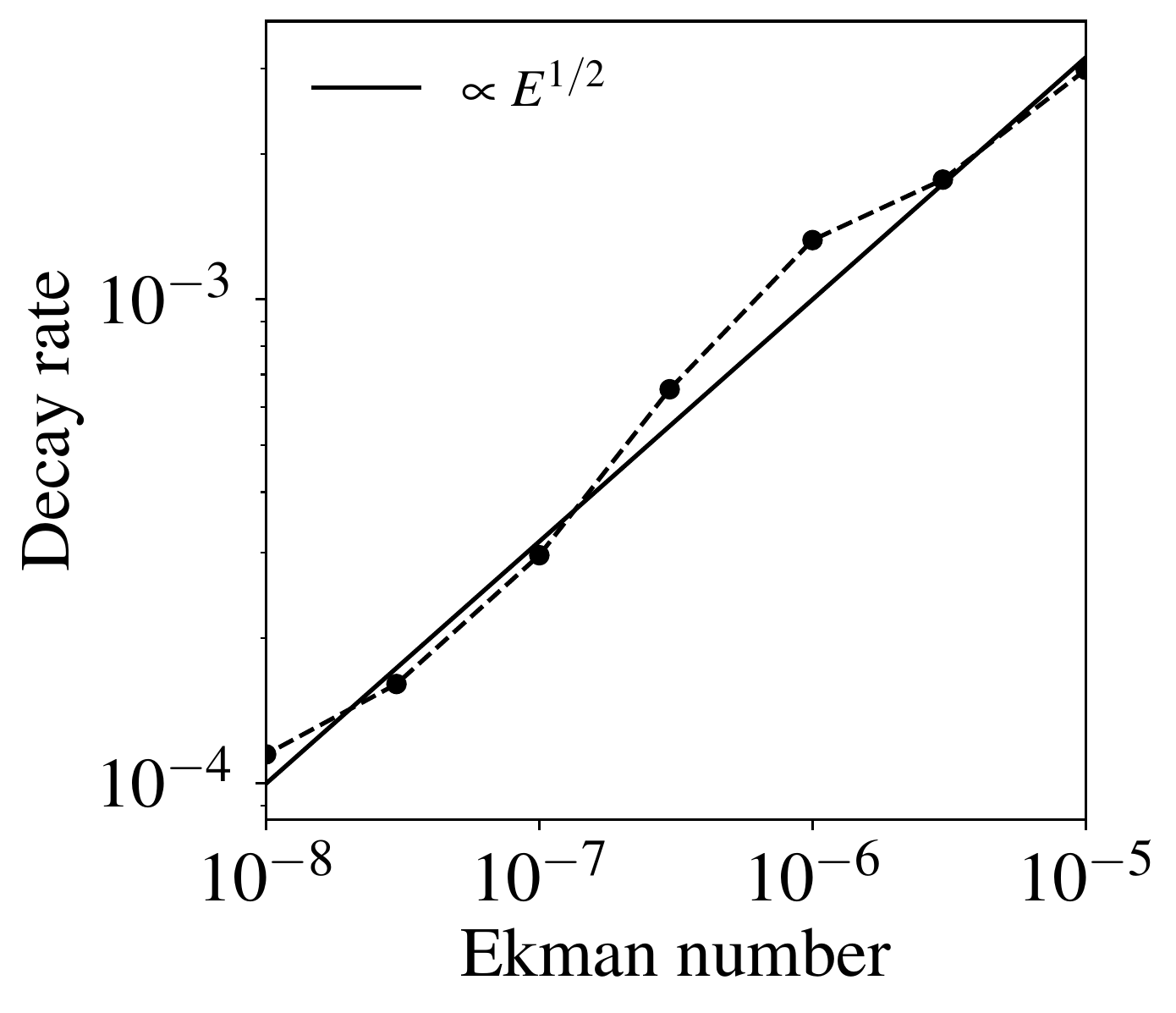}\\
(a) \hspace{7cm} (b)

\caption{(a) Dissipation rate for the forced flow at the resonant peak $\omega/\Omega=1.4406$ as a function of the Ekman number. (b) Decay rate of the least damped eigenmode around $\omega/\Omega=1.4406$ as a function of the Ekman number.}
\label{fig:DecayDissRate}
\end{figure*}

In order to understand the resonant behaviors, we compare tidally forced flows and eigenmodes at frequencies of peaks and troughs. \autoref{fig:ForcedEigen} shows two examples at $\omega/\Omega=1.15$ (corresponding to a trough) and  $\omega/\Omega=1.4406$ (corresponding to a peak)  respectively. The eigenmode is found through an iterative method described in \cite{Rieutord1997JFM} and represents the least damped mode around the forcing frequency. As the amplitude of an eigenmode is arbitrary, the velocity amplitude $|\mathbf{u}|$ is normalized to have the same maximum values as the corresponding forced flows. At $\omega/\Omega=1.15$, the eigenmode represents a clear wave attractor, which is also expected from the ray dynamics \citep{Ogilvie2009}. The forced flow at the same frequency shows that localized wave beams spawned from the critical latitude propagate along the characteristics and converge towards to the attractor of the eigenmode, but the forced flow and the eigenmode are considerably different as one can see from figure \ref{fig:ForcedEigen}(a). In contrast,  the forced flow and the eigenmode exhibit the same spatial structure at a peak frequency $\omega/\Omega=1.4406$ (figure \ref{fig:ForcedEigen}(b)), suggesting that the eigenmode is resonantly excited by the tidal forcing at this frequency. The resonance behavior at  $\omega/\Omega=1.4406$ is also evident in figure \ref{fig:DecayDissRate} (a), which shows that the dissipation rate of the tidally forced flow increases on reducing the Ekman number and scales approximately as $E^{-1/2}$. The corresponding eigenmode behaves similarly to normal modes with a damping rate that scales approximately as $E^{1/2}$. \cite{Rieutord2010} observed similar resonance behaviors and scalings for an axisymmetric mode at $\omega/\Omega=1.3196$ in a spherical shell with $r_i/r_o=0.35$, and noted that the resonance is difficult to explain in terms of ray dynamics.

\begin{figure*}
   \centering
\includegraphics[width=0.4 \textwidth]{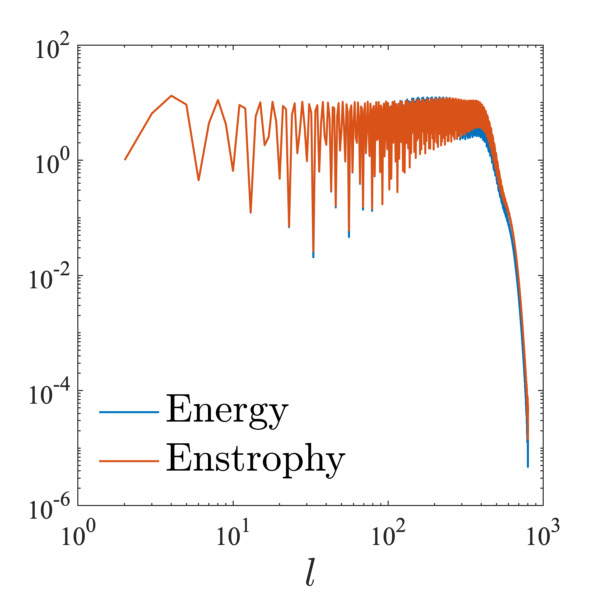}
\includegraphics[width=0.4 \textwidth]{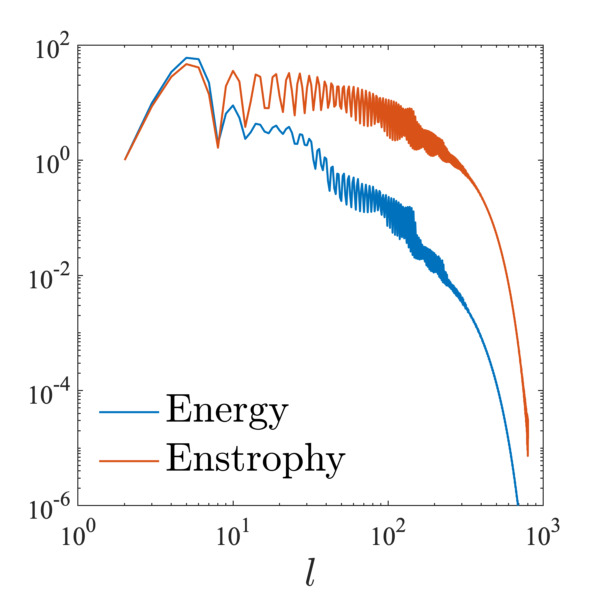}  \\
(a) \hspace{7cm} (b)
\caption{Energy and enstrophy spectra of the two eigenmodes in figure \ref{fig:ForcedEigen} as a function of the spherical harmonic degree $l$. (a) $\omega/\Omega=1.15$; (a) $\omega/\Omega=1.4406$.}
  \label{fig:Spectra}%
\end{figure*}

 Despite clear evidence of the resonance at $\omega/\Omega=1.4406$, the eigenmode features localized wave beams emanating from the critical latitude and propagating along the characteristics in the bulk. There should be some underlying mechanism responsible for the resonance and for intensifying the wave beams at the resonant frequencies.
\autoref{fig:Spectra} shows the spectra of energy ($|\mathbf u|^2$) and enstrophy ($|\mathbf {\nabla \times u}|^2$) as a function of the spherical harmonic degree $l$ for two eigenmodes in figure \ref{fig:ForcedEigen}. All spectra are normalized by the corresponding amplitude at the lowest degree $l=2$ as only the relative amplitude matters for the eigenmodes. We can see that the energy and enstrophy spectra have almost identical shapes for the attractor mode at $\omega/\Omega=1.15$, suggesting that the energy and dissipation are distributed on the same scales. However, there is a notable scale separation of the energy and enstrophy at around $l=8$ for the resonant eigenmode at $\omega/\Omega=1.4406$.  It seems that the energy is mainly contained in large scales but dissipation takes place at smaller scales for this mode. In fact, $l<8$ components contribute 60\% of the total energy but contain less than 7\% of the total enstrophy for the resonant eigenmode at $\omega/\Omega=1.4406$. This suggests the resonant mode contains significant large scale flows, which may hold the key for the resonant tidal response, though it is not obvious from figure \ref{fig:ForcedEigen} (b).

\begin{figure*}
 \centering
\includegraphics[width=0.32 \textwidth]{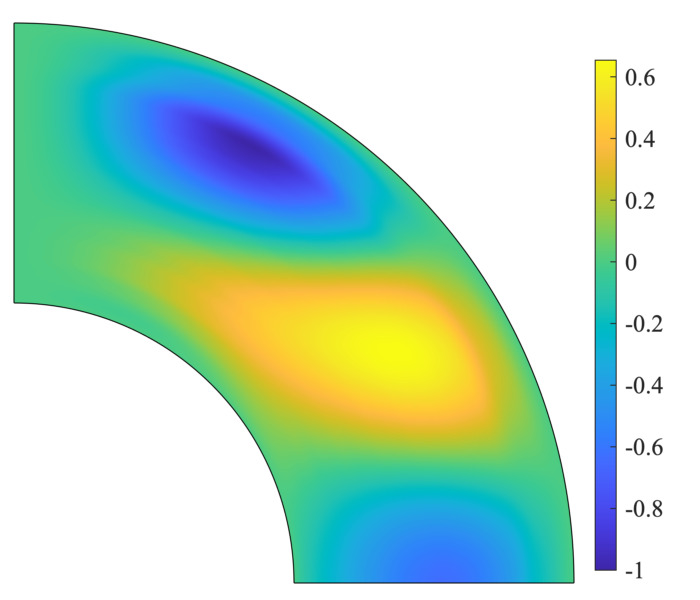}
\includegraphics[width=0.32 \textwidth]{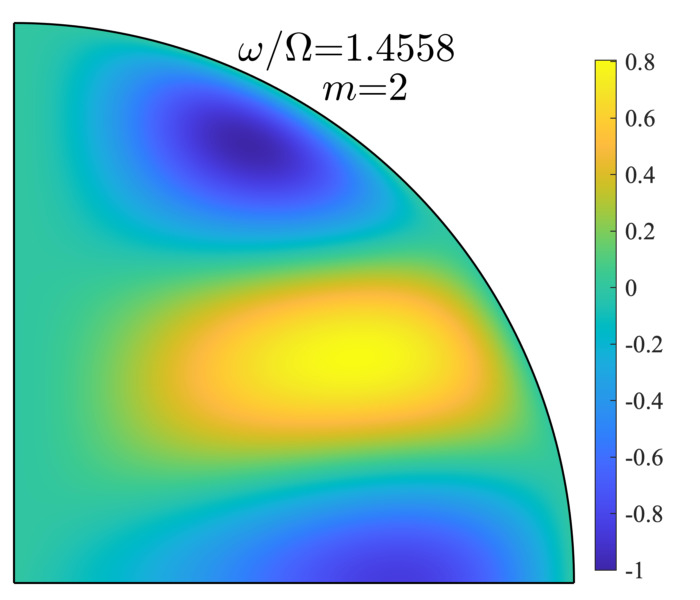}
\includegraphics[width=0.32 \textwidth]{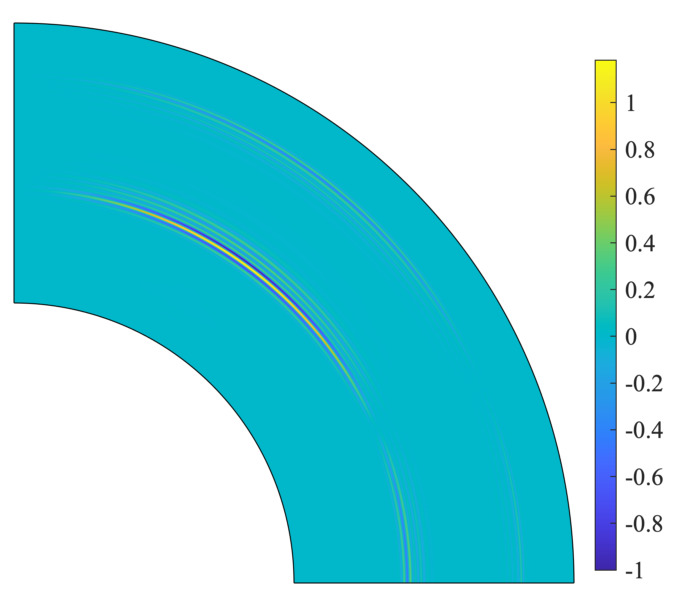} \\
(a) \hspace{5cm} (b) \hspace{5cm}  (c)
 \caption{Radial velocity in the meridional plane of eigenmodes. (a) Eigenmode at $\omega/\Omega=1.4406$ but including only contributions from $l< 8$. (b) Inviscid inertial mode in a full sphere at $\omega/\Omega=1.4558$ and $m=2$. (c) Eigenmode at $\omega/\Omega=1.15$ but including only contributions from $l<8$. }
 \label{fig:Modes}%
\end{figure*}

In order to reveal the large-scale flows of the eigenmode at  $\omega/\Omega=1.4406$,  figure \ref{fig:Modes}(a) shows the radial velocity of the eigenmode but including only spherical harmonics of $l<8$ (and $l\ge m=2$) as suggested by the spectra in figure \ref{fig:Spectra}(b). We see a smooth large-scale mode after the truncation, which is reminiscent of an inviscid inertial mode  in a full sphere with eigenfrequency $\omega/\Omega=1.4558$ and $m=2$  as shown in figure \ref{fig:Modes} (b). 
For comparison, we also show the radial velocity truncated to $l<8$ for the attractor mode at $\omega/\Omega=1.15$ in figure \ref{fig:Modes} (c), which just shows some local perturbations even after the truncation, suggesting that attractor modes do not have large-scale structure hidden beneath wave beams. This is the underlying difference between the resonant modes and attractor modes. 

In fact, for all of the major peaks in the dissipation curves (vertical dashed lines in figure \ref{fig:FreScan}), we find eigenmodes exhibiting large-scale smooth structures after truncation at appropriate $l$ as shown in figure A1 in the Appendix. The truncation for the large-scale flows is chosen based on the spectra of energy and enstrophy. For each mode, we also find a corresponding inviscid inertial mode in a full sphere with similar structure and nearby eigenfrequency as shown in figure A1.  We also show two modes for the radius ratios $r_i/r_o=0.35$ and  $r_i/r_o=0.7$ in  figure A2 in the Appendix.
It seems that resonant eigenmodes in a spherical shell with a moderate core size retain some features of inertial modes in a full sphere. {The large-scale mode ``leaks" by radiating a small-scale wave at the critical latitude, which propagates along the characteristics in the bulk. The picture is somewhat related to the wave scattering process described by \cite{Goodman2009ApJ}, but they assumed that the large-scale flow is  a plane wave and that the scattered wave breaks nonlinearly without subsequent reflections. We show that the large-scale wave is associated with eigenmodes of the system and therefore can account for the frequency-dependence of tidal dissipation.}

 However, we should note that the large-scale structures and the localized wave beams  together form inertial modes with a small viscosity in a spherical shell. The inviscid inertial wave problem remains singular and cannot be solved analytically in a shell. We separate two parts based on the spectra to illustrate that there are large-scale structures hidden beneath localized wave beams for the eigenmodes corresponding to resonant peaks. One would expect strong couplings between the large-scale tidal forcing (e.g. $l=m=2$) and the eigenmodes with hidden large scale flows, which can intensify the flux in wave beams emanating from the critical latitude and lead to enhanced tidal dissipation. This resolves a long-standing puzzle about the resonant peaks in the dissipation curves due to inertial waves.

\section{Conclusion}
Using a simplified model, we have shown that resonant tidal responses in the convective envelopes of rotating fluid bodies can be attributed to the excitation of  inertial modes which have large-scale structures hidden beneath localized wave beams. The hidden large-scale structures are revealed by analyzing the energy and enstrophy spectra of inertial modes in a spherical shell. We also note that these global smooth structures are reminiscent of inertial modes with similar eigenfrequencies in a full sphere. These particular eigenmodes are fundamentally different from the well-known attractor modes, which have been mistakenly linked to resonant tidal responses. Strong couplings  between the tidal forcing and inertial modes with hidden large-scale structures can explain the resonant peaks in the dissipation curves. Our results resolve a long-standing puzzle regarding the frequency dependence of tidal dissipation associated with inertial waves in convective envelopes.

In this study, we use a constant density model for the convective envelope, which is not realistic. However, it was found that inertial modes with smooth background density variations are similar to those with a constant density in a full sphere \citep{Wu2005ApJ}. If this remains true in a spherical shell, resonant inertial modes with hidden large-scale flows can give rise to non-negligible  gravitational perturbations considering background density variations. This is of particular importance for Jupiter and Saturn as resonant inertial modes in the convective envelopes can be potentially detected by high precision gravity measurements \citep{Idini2021PSJ}. However, this requires detailed calculations of inertial modes with more realistic models to quantitatively characterize possible detections. 

\acknowledgments
Y.L. was supported by the B-type Strategic Priority Program of the CAS (XDB41000000) and the pre-research project on Civil Aerospace Technologies of CNSA (D020308) and the Macau Foundation. This study was also supported by the Isaac Newton Trust in Cambridge when this work started.
Numerical calculations were performed on the Taiyi cluster supported by the Center for Computational Science and Engineering of Southern University of Science and Technology.


\bibliography{tides}
\bibliographystyle{aasjournal}

\section*{Appendix}
In this Appendix, we show more examples of eigenmodes having large-scale structures hidden beneath localized wave beams. Figure A1 shows four modes corresponding peaks in the dissipation curves in Figure 1. Figure A2 shows two modes with the radius ratio of 0.35 and 0.7 respectively.

\begin{figure*}[!h]
\centering
\includegraphics[width=0.9 \textwidth]{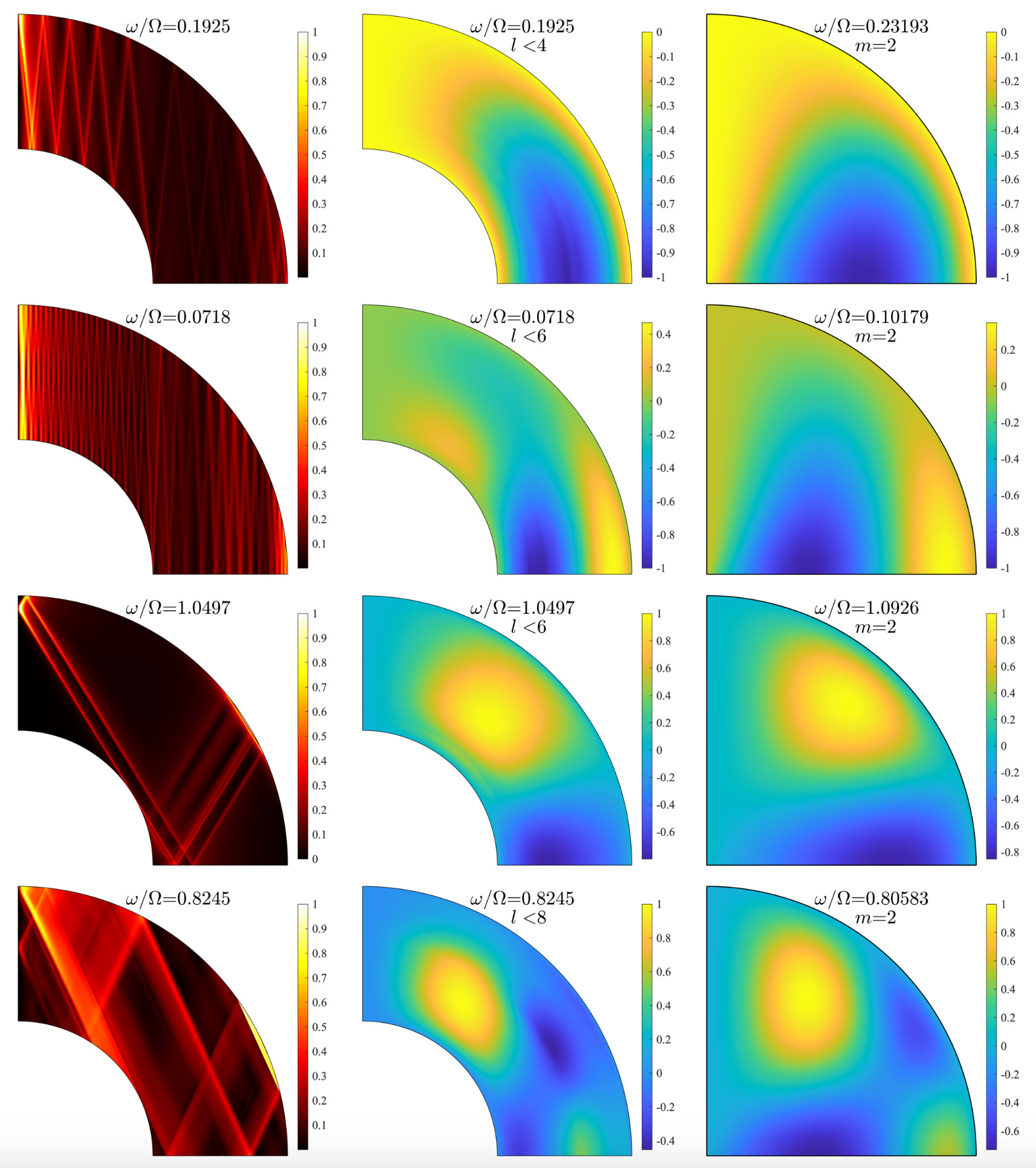}\\
\textbf{Figure A1}. {Eigenmodes  at frequencies corresponding to several peaks in the dissipation curves. Left column: velocity amplitude $|\mathbf{u}|$; Middle column: radial velocity truncated at low degree of spherical harmonics; Right column: inviscid inertial modes in a full sphere with similar frequency and structure as in the middle column. $E=10^{-8}$ for modes in a spherical shell.}
\end{figure*}

\begin{figure*}[!h]
\centering
\includegraphics[width=0.9 \textwidth]{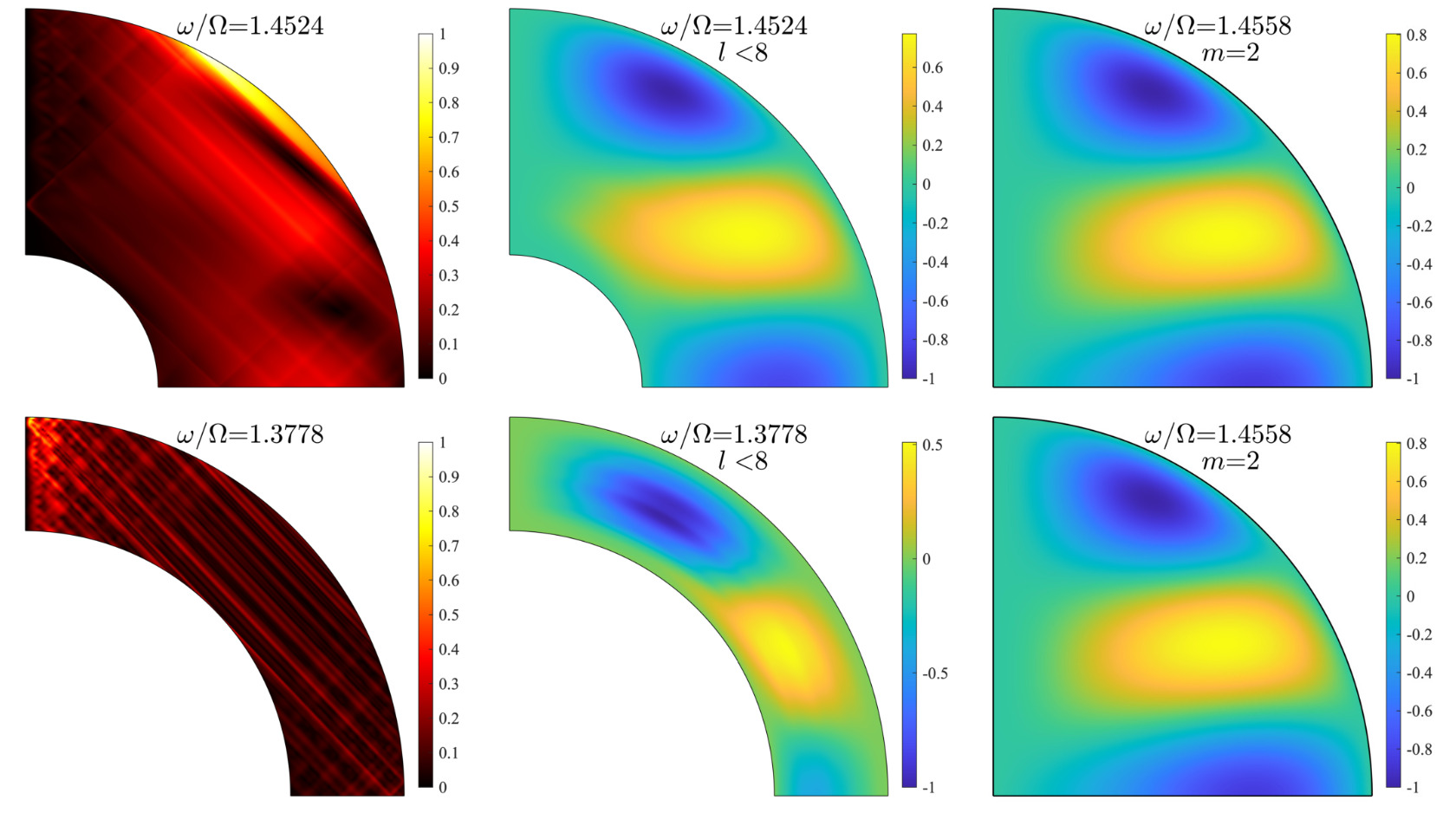}\\
\textbf{Figure A2}. {As for figure A1 but for two modes with the radius ratio $r_i/r_o=0.35$ (top panel) and $r_i/r_o=0.7$ (bottom panel) .}
\end{figure*}

\end{document}